\documentclass[hyper]{JHEP} 
\usepackage{epsfig}
\newcommand\fverb{\setbox\pippobox=\hbox\bgroup\verb}
\newcommand\fverbdo{\egroup\medskip\noindent%
            \fbox{\unhbox\pippobox}\ }

\newcommand\fverbit{\egroup\item[\fbox{\unhbox\pippobox}]}
\newbox\pippobox
\title{Spiky Strings on NS5-branes}
\author{Sagar Biswas\\
Department of Phyisics and Meteorology, \\
Indian Institute of technology Kharagpur,\\
Kharagpur-721 302, INDIA \\
E-mail: \email{biswas.sagar09iitkgp@gmail.com}}
\author{Kamal L. Panigrahi\\
Department of Physics and Meteorology \&
Centre for Theoretical studies,\\
Indian Institute of Technology Kharagpur,\\
Kharagpur-721302, INDIA \\
E-mail: \email{panigrahi@phy.iitkgp.ernet.in}} \abstract{We study
rigidly rotating strings in the near horizon geometry of a stack
of Neveu-Schwarz (NS) 5-branes. We solve the Nambu-Goto action of
the fundamental string in the presence of a NS-NS two form
$(B_{\mu\nu})$ and find out limiting cases corresponding to magnon
and spike like solutions.} 

\begin{document}
\section{Introduction} The conjectured AdS/CFT duality \cite{Maldacena:1997re}
has passed through various nontrivial tests in the past by
analyzing the spectrum of quantum string states on AdS background
and the spectrum of the anomalous dimensions of the $N=4$ gauge
theory operators in {\it planar limit}. Especially it has been
noticed that in the semiclassical approximation the theory becomes
{\it integrable} on the both sides of the duality. Though finding
out the full spectrum of string theory in AdS background is
difficult, it has been observed that in certain limits, for
example in large angular momentum sector, the theory
\cite{Berenstein:2002jq} is more tractable. In this region, one
can use the semiclassical approximations to find the string
spectrum as well \cite{Gubser:2002tv}. The semiclassical string
states in the gravity side has been used to look for suitable
gauge theory operators on the boundary, in establishing the
duality. In this connection, Hofman and Maldacena (HM)
\cite{Hofman:2006xt} considered a special limit \footnote{The
Hofman-Maldacena limit: $J\rightarrow \infty, \lambda = {\rm
fixed}, p= {\rm fixed}, E-J = {\rm fixed}$} where the problem of
determining the spectrum of both sides becomes rather simple. The
spectrum consists of an elementary excitation known as magnon
which propagates with a conserved momentum $p$ along the spin
chain. Further, a general class of rotating string solution in
$AdS_5$ is the spiky string which describes the higher twist
operators from dual field theory view point. Giant magnons can be
thought of as a special limit of such spiky strings with shorter
wavelength. Recently there has been a lot of work devoted for the
understanding of the giant magnon and spiky string solutions in
various backgrounds, see for example in \cite{Dorey:2006dq}-
\cite{Kluson:2008gf}. There has also been numerous papers devoted
for understanding the finite size corrections on these solutions,
see \cite{Arutyunov:2006gs,Astolfi:2007uz} and references therein.

In the present paper we generalize the discussion of spiky string
in the near horizon geometry of a stack of NS5-branes
\cite{Kluson:2007qu}. In string theory NS5-branes are interesting
because in the near horizon limit the theory on the worldvolume
correspond to a nonlocal field theory, namely the little string
theory (LST) \footnote{For review, see
\cite{Aharony:1999ks,Kutasov:2001uf}.}. Though the LST has not
been understood properly until now, it seems a good exercise to
analyze the solution in various limits and find out some operators
in dual field theory if possible. The near horizon NS5-brane world
sheet theory is exactly solvable, so from the bulk theory view
point the theory is integrable. A little is known about the
boundary theory, hence from that prospective it is rather hard to
make definite statements about the exact nature of the theory.
However it is interesting to study semiclassical spiky like string
solution in the linear dilaton background and look for suitable
operators in the boundary theory. Indeed such an attempt was made
in \cite{Kluson:2007qu} and the giant magnon solution with one
angular momentum was found out. However, in deriving the same one
had to look for limits where the background three form flux that
supports the NS5-brane background vanishes. We would like to
generalize that to a non-vanishing three form flux in the present
paper. In deriving so, we find out a general class of rotating
string solutions. In different limit, we found out solutions
corresponding to giant magnon and spiky like string. The
dispersion relation obtained is slightly different from that of
the usual relation among various charges typically known for the
case of two and three-spheres. The difference comes about because
the presence of the background three form flux in the supergravity
background which couples to the Fundamental string. The results
can be compared with that of the solutions in the presence of
background B$_{\rm{NS-NS}}$ fields.

Rest of the paper is organized as follows. In the section-2, we
write down the Nambu-Goto action for the rigidly rotating string
in the near horizon limit of the NS5-brane background. We find
various conserved charges, namely the total energy $E$, momenta
along two angular direction $(\phi_1, \phi_2)$ of the 3- sphere in
the transverse direction of NS 5-brane, momenta along the five
longitudinal directions of NS5-branes, $y^i$, and another
conserved charge $D$ related to the radial motion of the string.
In section-3, we analyze two different limits of the solution
which correspond to magnon and single spike string. We adopt a
particular regularization to find out the dispersion relation
among various charges. Finally in section-4 we conclude with some
discussions.
\section{Rigidly rotating Strings in NS5-brane Background}
In this section we study rigidly rotating string in the near
horizon limit of a stack of NS5-branes\footnote{some aspects of
string dynamics in NS5-brane background is studied for example in
\cite{Madrigal:2009nf}}. The classical solution of $N$ NS5-brane
is given by the following form of metric, NS-NS two form
$(B_{\mu\nu})$ field and the dilaton,
\begin{eqnarray}
ds^2 &=& -d\tilde{t}^2 + d \tilde{y}^2_{i} + H(r)\left(dr^2 + r^2(d\theta^{2} + \sin^2\theta d\phi_1^2 +\cos^2\theta d\phi_2^2)\right) \cr \nonumber \\
e^{2(\phi-\phi_0)}&=& H(r), \>\> B = 2 N \sin^2\theta d\phi_1
\wedge d\phi_2, \>\> H(r) = 1+\frac{Nl{_s}{^2}}{r^2}
\end{eqnarray}
where $y^i$, $i=5, 6,7,8,9$ label the  world-volume directions of
NS5-brane, and $H(r)$ is the harmonic function in the transverse
space of the NS5-branes, $l_s$ is the string length. In the near
horizon limit,  $r\rightarrow 0$, one can ignore 1 in $H(r)$, and
the solution would look like (defining $\tilde{t}=\sqrt{N}l{_s} t
, \tilde{y_i} = \sqrt{N}l_s y_i$)
\begin{eqnarray}
ds^2  &=& Nl{_s}{^2}\left(-dt^2+d\theta^{2} + \sin^2\theta d\phi_1^2 +\cos^2\theta d\phi_2^2
+\frac{dr^2}{r^2} + dy_{i}^{2}\right), \cr \nonumber \\
e^{2(\phi-\phi_0)}&=& \frac{Nl{_s}{^2}}{r^2}, \>\> B = 2 N \sin^2\theta d\phi_1 \wedge d\phi_2,
\end{eqnarray}
To proceed further it is convenient to introduce the variable
$\rho$ that is related to $r$ as
\begin{eqnarray}
\rho=\ln\left(\frac{r}{\sqrt{Nl{_s}{^2}}}\right)
\end{eqnarray}
so that the metric becomes,
\begin{equation}
ds^2  =Nl{_s}{^2}\left(-dt^2+d\theta^{2} + \sin^2\theta d\phi_1^2
+\cos^2\theta d\phi_2^2 +d\rho^{2} + dy_{5}^{2}\right) \ ,
\end{equation}
with the two form B$_{\phi_1 \phi_2}$ remaining the same. We wish
to study rigidly rotating string in this background. The
Nambu-Goto action of the string in the presence of a background B
field is written as\footnote{in the classical approximation (large
`t Hooft coupling) we can ignore the dilaton coupling.}
\begin{eqnarray}
S = - \frac{\sqrt{\lambda}}{2\pi}\int{d^2 \sigma
\left[\sqrt{\tilde{g}} + \epsilon^{\alpha\beta} B_{\mu\nu}
\partial_{\alpha} X^{\mu} \partial_{\beta} X^{\nu}\right]} \ ,
\end{eqnarray}
where
\begin{eqnarray}
\tilde{g}_{\alpha\beta} = G_{\mu\nu} \partial_{\alpha}X^{\mu}
\partial_{\beta}X^{\nu}
\end{eqnarray}
is the induced metric on the string world sheet and the pre factor $\sqrt{\lambda}=N$
is the `t Hooft coupling constant.
The Lagrangian is given by
\begin{eqnarray}
&&\mathcal{L} = -\Big(\Big[{(-\dot{t} {t'} + \dot{\theta}{\theta'} + \sin^2\theta \dot{\phi_1}{\phi'}_1 + \cos^2\theta \dot{\phi_2}{\phi'}_2 + \dot{\rho}{\rho'} + \dot{y_i}{y'}_i)}^2  \nonumber \\
&&- (-\dot{t}^2 + \dot{\theta}{^2} + \sin^2\theta \dot{\phi_1}{^2}
+ \cos^2\theta \dot{\phi_2}^2 + \dot{\rho}^2 + \dot{y_i}^2)
(-{t'}^2 + {\theta'}{^2} + \sin^2\theta {\phi'}^2_1 + \cos^2\theta
{\phi'}^2_2
+ {\rho'}^2 + {y'}^2_i)\Big]^{1/2}  \nonumber \\
&&+ 2 \sin^2\theta \left(\dot{\phi_1}{\phi'}_2 - \dot{\phi_2}{\phi'}_1\right)\Big) \nonumber \\
\end{eqnarray}
For further analysis we choose the following ansatz
\begin{eqnarray}
t=\kappa \tau ; \> \theta=\theta(\sigma); \> \phi_1=\nu_1 \tau + \sigma ; \> \phi_2=\nu_2 \tau + \phi(\sigma) ; \> \rho= m \tau ; \> y_i= v_i\tau
\end{eqnarray}
The Euler-Lagrangian equations derived from the above action are given by
\begin{eqnarray}
\partial_{\sigma}\frac{\partial\mathcal{L}}{\partial t'} + \partial_{\tau}\frac{\partial\mathcal{L}}{\partial {\dot{t}}} = \frac{\partial \mathcal{L}}{\partial t}
\label{EL-t}
\end{eqnarray}
\begin{eqnarray}
\partial_{\sigma}\frac{\partial\mathcal{L}}{\partial {\rho}'} + \partial_{\tau}\frac{\partial\mathcal{L}}{\partial {\dot{\rho}}} = \frac{\partial \mathcal{L}}{\partial \rho}
\label{EL-rho}
\end{eqnarray}
\begin{eqnarray}
\partial_{\sigma}\frac{\partial\mathcal{L}}{\partial {\phi}'_1} + \partial_{\tau}\frac{\partial\mathcal{L}}{\partial {\dot{\phi_1}}} = \frac{\partial \mathcal{L}}{\partial \phi_1}
\label{EL-p1}
\end{eqnarray}
\begin{eqnarray}
\partial_{\sigma}\frac{\partial\mathcal{L}}{\partial {\phi}'_2} + \partial_{\tau}\frac{\partial\mathcal{L}}{\partial {\dot{\phi_2}}} = \frac{\partial \mathcal{L}}{\partial \phi_2}\label{EL-p2}
\end{eqnarray}
\begin{eqnarray}
\partial_{\sigma}\frac{\partial\mathcal{L}}{\partial {y}'_i} + \partial_{\tau}\frac{\partial\mathcal{L}}{\partial {\dot{y_i}}} = \frac{\partial \mathcal{L}}{\partial y_i}\label{EL-y}
\end{eqnarray}
First solving (\ref{EL-t}) and (\ref{EL-p1}) we get
\begin{eqnarray}
\phi'=\frac{\sin^2\theta(\alpha^2C_1-\kappa\nu_1C_2-\nu_2^2C_1\cos^2\theta+2\kappa \nu_1\nu_2\sin^2\theta)}{\nu_2\cos^2\theta(\kappa C_2-C_1\nu_1\sin^2\theta-2\kappa \nu_2 \sin^2\theta)}
\label{diff-phi}
\end{eqnarray}
and for $\theta$, we get
\begin{eqnarray}
{\theta'}^2=\frac{{(\kappa^2-C_1^2)(\nu_1\sin^2\theta + \phi^{\prime}\nu_2\cos^2\theta)}^2}{C_1^2(\alpha^2-\nu_1^2\sin^2\theta -\nu_2^2\cos^2\theta)}-\sin^2\theta-{\phi'}^2\cos^2\theta
\label{diff-theta}
\end{eqnarray}
Note that writing the above two equations, we have made use of the
solutions obtained from (\ref{EL-t}) and (\ref{EL-p1}). Further,
$C_1$ and $C_2$ are integration constants and $\alpha =
\sqrt{\kappa^2 - m^2 - v_i v^i}$. Note also that by looking on the
background, the Lagrangian is invariant under the following
translation
\begin{eqnarray}
t'(\tau, \sigma) &=& t (\tau, \sigma) + \epsilon_t, \>\>\> {\phi}'_1(\tau, \sigma) = \phi_1 (\tau, \sigma) + \epsilon_{\phi_1}, \>\>\> {\phi}'_2 (\tau, \sigma) = \phi_2 (\tau, \sigma) + \epsilon_{\phi_2},
\nonumber \\
\rho'(\tau, \sigma) &=& \rho (\tau, \sigma) + \epsilon_{\rho}, \>\>\> {y}'_i(\tau, \sigma) = y_i (\tau, \sigma) + \epsilon_{y_{i}} \ ,
\end{eqnarray}
where $\epsilon_{t}, \epsilon_{\phi_1}, \epsilon_{\phi_2}, \epsilon_{\rho}, \epsilon_{y_i}$ are constants.
Hence the rigidly rotating string has a number of conserved charges, namely the total energy $(E)$,
angular momenta $J_1$, and $J_2$, and $D$, which comes from the translation of $\rho$, and five conserved
charges, $ P_{y_i}$ coming from the translational invariance of the metric along $y^i$-directions. We
will compute them and find out relation between them for different limits where one can compute them
easily.
\section{Limiting Cases}
In this section, we will analyze two limiting cases corresponding
to the usual single spike and giant magnon like solutions
respectively. First we will look for the spike solution.
\subsection{single spike} We choose the constants of motion appropriately, such that
$\theta' \rightarrow 0$ as $\theta \rightarrow \frac{\pi}{2}$, the values of constants we get
are \begin{eqnarray}
C_2=\alpha +2\nu_2, \>\>\>\> C_1\alpha=\kappa \nu_1
\end{eqnarray}
With these constants, the differential equations for $\theta$ and $\phi$ now becomes
\begin{eqnarray}
\phi'=-\frac{\nu_1(\nu_2+2\alpha)\sin^2\theta}{\alpha^2+2\nu_2\alpha \cos^2\theta -\nu_1^2\sin^2\theta}
\end{eqnarray}
and
\begin{eqnarray}
\theta'=\frac{\alpha\sqrt{3(\nu_2^2-\nu_1^2)}\sin \theta \cos \theta \sqrt{\sin^2\theta-\sin^2\theta_0}}{\alpha^2+2\nu_2\alpha \cos^2\theta -\nu_1^2\sin^2\theta}
\end{eqnarray}
where
\begin{eqnarray}
\sin \theta_0=\frac{\alpha+2\nu_2}{\sqrt{3(\nu_2^2-\nu_1^2)}}
\end{eqnarray}
Now we compute the conserved quantities by the following
relations. The energy is (define $T =
\frac{\sqrt{\lambda}}{{2\pi}}$)
\begin{eqnarray}
 E = -2T\int_{\theta_0}^{\theta_1}\frac{d\theta}{\theta'}\frac{\partial \mathcal{L}}{\partial \dot{t}}=\frac{2T\kappa (\alpha ^2-\nu_1^2)}{\alpha^2 \sqrt{3(\nu_2^2-\nu_1^2)}}\int_{\theta_0}^{\frac{\pi}{2}}\frac{\sin \theta d\theta}{\cos \theta \sqrt{\sin^2\theta-\sin^2\theta_0}}
\end{eqnarray}
Angular momenta
\begin{equation}
J_1 = 2T\int_{\theta_0}^{\theta_1}\frac{d\theta}{\theta^{\prime}}\frac{\partial \mathcal{L}}{\partial \dot{\phi_1}}=-\frac{2T\nu_1}{\alpha \sqrt{3(\nu_2^2-\nu_1^2)}}\int_{\theta_0}^{\frac{\pi}{2}}\frac{\sin \theta[2(2\alpha +\nu_2)-3\alpha\cos^2\theta] d\theta}{\cos \theta \sqrt{\sin^2\theta-\sin^2\theta_0}}
\end{equation}
and
\begin{equation}
J_2 =
2T\int_{\theta_0}^{\theta_1}\frac{d\theta}{\theta^{\prime}}\frac{\partial
\mathcal{L}} {\partial \dot{\phi_2}}=\frac{2T}{\alpha
\sqrt{3(\nu_2^2-\nu_1^2)}}\int_{\theta_0}^{\frac{\pi}{2}}
\frac{\sin \theta[2(\nu_1^2 -\alpha^2)-3\alpha \nu_2 \cos^2\theta]
d\theta}{\cos \theta \sqrt{\sin^2\theta-\sin^2\theta_0}}
\end{equation}
Further we get more conserved quantities defined by
\begin{equation}
D=2T\int_{\theta_0}^{\theta_1}\frac{d\theta}{\theta^{\prime}}\frac{\partial \mathcal{L}}{\partial \dot{\rho}}=\frac{2Tm(\alpha ^2-\nu_1^2)}{\alpha^2 \sqrt{3(\nu_2^2-\nu_1^2)}}\int_{\theta_0}^{\frac{\pi}{2}}\frac{\sin \theta d\theta}{\cos \theta \sqrt{\sin^2\theta-\sin^2\theta_0}}
\end{equation}
and
\begin{equation}
P_{y_i}=2T\int_{\theta_0}^{\theta_1}\frac{d\theta}{\theta^{\prime}}\frac{\partial \mathcal{L}}{\partial \dot{y_i}}=\frac{2Tv(\alpha ^2-\nu_1^2)}{\alpha^2 \sqrt{3(\nu_2^2-\nu_1^2)}}\int_{\theta_0}^{\frac{\pi}{2}}\frac{\sin \theta d\theta}{\cos \theta \sqrt{\sin^2\theta-\sin^2\theta_0}} \ .
\end{equation}
Further, the difference in the angle between two end points of the string is
\begin{equation}
 \Delta\phi = 2\int_{\theta_0}^{\theta_1}\frac{d\theta}{\theta^{\prime}}=-\frac{2}{\alpha\sqrt{3(\nu_2^2-\nu_1^2)}}\int_{\theta_0}^{\frac{\pi}{2}}\frac{\nu_1^2\sin^2\theta-2\nu_2\alpha\cos^2\theta-\alpha^2}{\cos \theta \sin \theta \sqrt{\sin^2\theta-\sin^2\theta_0}}
\end{equation}
Here we can see that all conserved quantities diverges, but the quantity
\begin{eqnarray}
\left(T\Delta \phi-\sqrt{E^2-D^2-P^2_{y_i}}\right)=2T \left(\frac{\pi}{2}-\theta_0\right)
\end{eqnarray}
is finite.

Further, one can regularize $J_1$ and $J_2$ by using the expression for $E$ as follows
\begin{equation}
\tilde{J_1} = J_1-\frac{2\nu_1(\nu_2+2\alpha)}{\nu_1^2-\alpha^2}\sqrt{E^2-D^2-P^2_{y_i}}
= \frac{6T \nu_1}{\sqrt{3(\nu_2^2-\nu_1^2)}}\cos\theta_0
\end{equation}
\begin{equation}
\tilde{J_2} = J_2 + 2\sqrt{E^2-D^2-P^2_{y_i}}=-\frac{6T\nu_2}{\sqrt{3(\nu_2^2-\nu_1^2)}}\cos\theta_0
\end{equation}
and they hold the following dispersion relation
\begin{equation}
\tilde{J_1}=\sqrt{\tilde{J_2^2}- \frac{3\lambda}{\pi^2}
\sin^2(\frac{\pi}{2} - \theta_0)} \ .
\end{equation}
As pointed out earlier the above relation should be compared with
the dispersion relation typically known for the case of S$^3$ with
the presence of B$_{\mu\nu}$ field $(b)$, \cite{Chen:2008vc} i.e.
\begin{eqnarray}
\tilde{J_1} = \sqrt{{\tilde J}^2_2 + \frac{\lambda}{{\pi^2}}
(1-b^2) \sin^2(\frac{\pi}{2} - \theta_0)}
\end{eqnarray}

\subsection{Giant Magnon} Now we consider the opposite limit, where the constants are
\begin{eqnarray}
\alpha=\nu_1, \>\>\>\>
\kappa C_2=\nu_1C_1 + 2\kappa \nu_2 \ .
\end{eqnarray}
Substituting these we get the following differential equations for $\phi$ and $\theta$,
\begin{equation}
\phi' = -\frac{\sin^2\theta(\nu_2C_1 + 2\kappa \nu_1)}{\cos^2\theta(\nu_1C_1 + 2\kappa \nu_2)}
\end{equation}
and
\begin{equation}
\theta' = \frac{\sin\theta \sqrt{\sin^2\theta-\sin^2\theta_1}}{\cos\theta \sin \theta_1}
\end{equation}
where
\begin{eqnarray}
\sin \theta_1= \frac{\nu_1 C_1 + 2\kappa \nu_2}{\kappa \sqrt{3(\nu_2^2-\nu_1^2)}}
\end{eqnarray}
Now the conserved charges become, the energy $(E)$ as
\begin{eqnarray}
 E =\frac{2T(\kappa ^2-C_1^2)}{\kappa \sqrt{3(\nu_2^2-\nu_1^2)}}\int_{\theta_1}^{\frac{\pi}{2}}\frac{\sin \theta d\theta}{\cos \theta \sqrt{\sin^2\theta-\sin^2\theta_1}}
\end{eqnarray}
Angular momenta along $\phi_1$ and $\phi_2$, respectively, are
\begin{eqnarray}
J_1 =
- \frac{2T}{\kappa^2
\sqrt{3(\nu_2^2-\nu_1^2)}}\int_{\theta_1}^{\frac{\pi}{2}}\frac{\sin
\theta[3\nu_1 \kappa^2\cos^2\theta -(\nu_1 C_1^2 + 2\kappa \nu_2
C_1 +3\nu_1\kappa^2)] d\theta}{\cos \theta
\sqrt{\sin^2\theta-\sin^2\theta_1}}
\end{eqnarray}
and
\begin{eqnarray}
J_2
=-\frac{6T\nu_2}{\sqrt{3(\nu_2^2-\nu_1^2)}}\cos\theta_1
\end{eqnarray}
Further conserved charges, derived from the translation along $\rho$ and along $y_i$ are
\begin{eqnarray}
D = \frac{2Tm(\kappa ^2-C_1^2)}{\kappa^2 \sqrt{3(\nu_2^2-\nu_1^2)}}\int_{\theta_1}^{\frac{\pi}{2}}\frac{\sin \theta d\theta}{\cos \theta \sqrt{\sin^2\theta-\sin^2\theta_1}}
\end{eqnarray}
and
\begin{eqnarray}
P_{y_i} = \frac{2Tv(\kappa ^2-C_1^2)}{\kappa^2 \sqrt{3(\nu_2^2-\nu_1^2)}}\int_{\theta_1}^{\frac{\pi}{2}}\frac{\sin \theta d\theta}{\cos \theta \sqrt{\sin^2\theta-\sin^2\theta_1}}
\end{eqnarray}
also the difference in the angle between two ends of the string is
\begin{eqnarray}
 \Delta\phi = 2\int_{\theta_0}^{\theta_1}\frac{d\theta}{\theta^{\prime}}= \pi- 2\theta_1 = p
\end{eqnarray}
Thus we see that here $J_2$ and $\Delta \phi$ are finite and rest
of the quantities diverge.
We can further rescale the expression for energy $\sqrt{E^2- D^2 -
P^2_{y_{i}}}$ as
\begin{eqnarray}
\tilde{E}=\frac{\nu_1C_1^2 + 2\kappa \nu_2 C_1 +
3\nu_1\kappa^2}{\alpha(\kappa^2-c_1^2)}\sqrt{E^2-D^2-P^2_{{y_i}}}
\end{eqnarray}
%
and then obtain the following dispersion relation
%


\begin{eqnarray}
\tilde{E}-{J}_1=\sqrt{{J^2_2-\frac{3\lambda}{\pi^2}\sin^2\frac{p}{2}}}
\ .
\end{eqnarray}
Once again this dispersion relation among energy and angular
momenta should be compared with the relation obtained in case of
S$^3$ in the presence of a NS-NS $B$ field\cite{Chen:2008vc}. Note
that in the above relation, the presence of the charge $D$
($\tilde{E}$ contains $D$) does not imply any new physical
interpretation of the dispersion relation. It simply reflects the
fact that the motion of the string along the radial direction in
the near horizon limit of NS5-brane background is free.
\section{Conclusions} In this paper we have studied rigidly
rotating strings in NS5-brane background. We have found out a
general class of solutions and have studied some limiting cases,
which correspond to spiky strings. In the process we have used a
particular type of regularization to derive the relationship among
various conserved charges. We have found out the dispersion
relation that is different from the usual dispersion relation of
the giant magnon and spiky strings on S$^3$
\cite{Ishizeki:2007we}. This stems from the fact that the F-string
couples to the NS-NS $B$-field that appears in the supergravity
background of a stack of NS5-branes. Hence the solutions of the
equations of motion for the $F$-string knows about it. The present
study can be extended in various directions. First, the finite
size corrections can be studied easily by following
\cite{Arutyunov:2006gs}. Further it would be challenging to study
the boundary theory operators of the solutions presented here.

\end{document}